# Macrogeneration and Automata Libraries
# For COSMA design environment

by
Wiktor B. Daszczuk [1]

**Summary**

In ICS, WUT a COSMA design environment is being developed. COSMA is based on Concurrent State Machines (CSM) formalism of system specification. It contains a graphical tool for system design, various tools for the analysis (including a temporal model checker), simulator and code generator. In many projects, some common susbsystems take place. This concerns both complicated modules and simple counters. In the report, a concept of macrogeneration technique for building of libraries of automata is presented. The new technique will support a compactness of projects and reusability of modules.

June 2003

---

[1] wbd@ii.pw.edu.pl

# INTRODUCTION

In the modeling of complex systems, many elements are repeatedly used for many times using the same or almost the same design. In a programming language they are called procedures (if one entry point and one exit point are applied) or classes (with more sophisticated structure). However, it is hard to organize such reusability in graphical specification language, such as used in COSMA modeling environment for CSM specifications of concurrent systems [Dasz01].

I will propose a method that is similar to macro-generation techniques. It provides two general features for the designer:
- using fragments of specification in various locations in the design (such as procedures or classes),

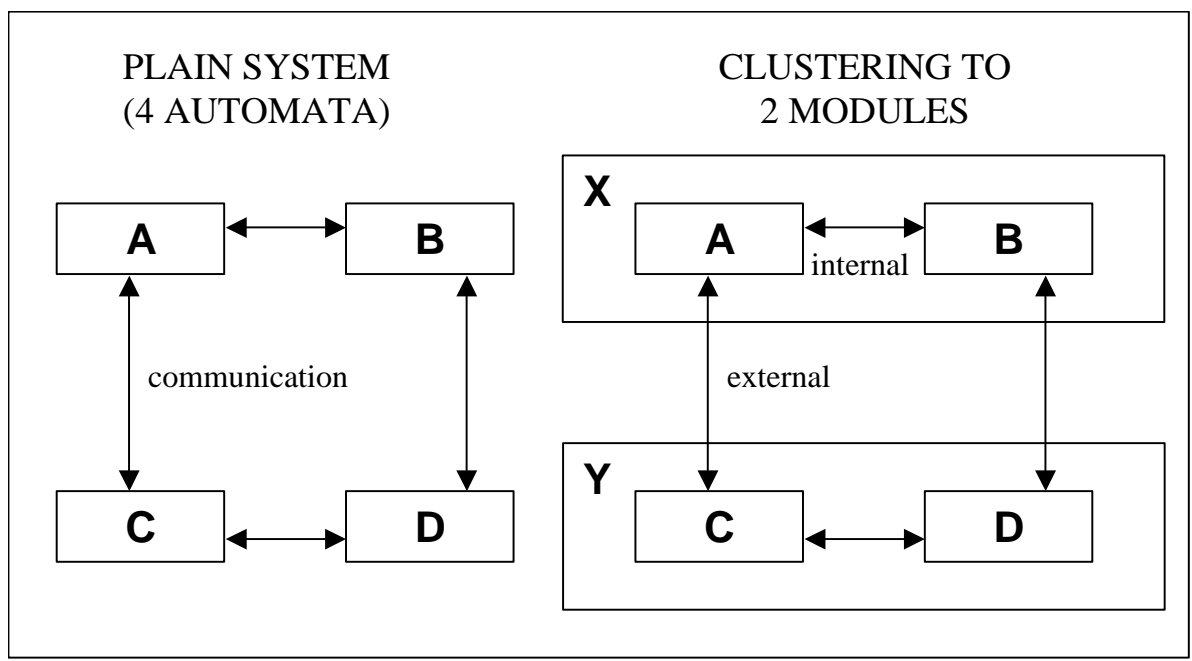

**Fig. 1 Clustering a structure of the system into modules**

- building libraries of reusable modules with published interface for use in many designs.

The most important feature of such organized reusability is that library modules may be verified using model checking facility in COSMA design environment [Cwww]. Thus, when using modules the designer need not verify the features that are published as proved.

In programming languages there is clear distinction between the manner of "external" communication (methods, parameters) from "internal" communication (variables, synchronization primitives). There is no such distinction in modeling formalism based on automata and signals passed between them [Mieś94, Mieś03]. Therefore we must impose some external structure onto the system of concurrent CSM automata:

- the set of all automata should be divided into subsets conforming separate modules in the system,
- signals passed between automata of a module will be treated as "internal" communication, while signals passed between automata belonging to separate modules will be treated as "external" communication.

Example of such a system (presented as block diagram, in which nodes are component automata and arcs denote communication between them) if shown in Fig. 1.

But in such structure, every module is self-contained and is used only once. We need also a mechanism for reusability, i.e. a kind of formal/actual parameter passing.

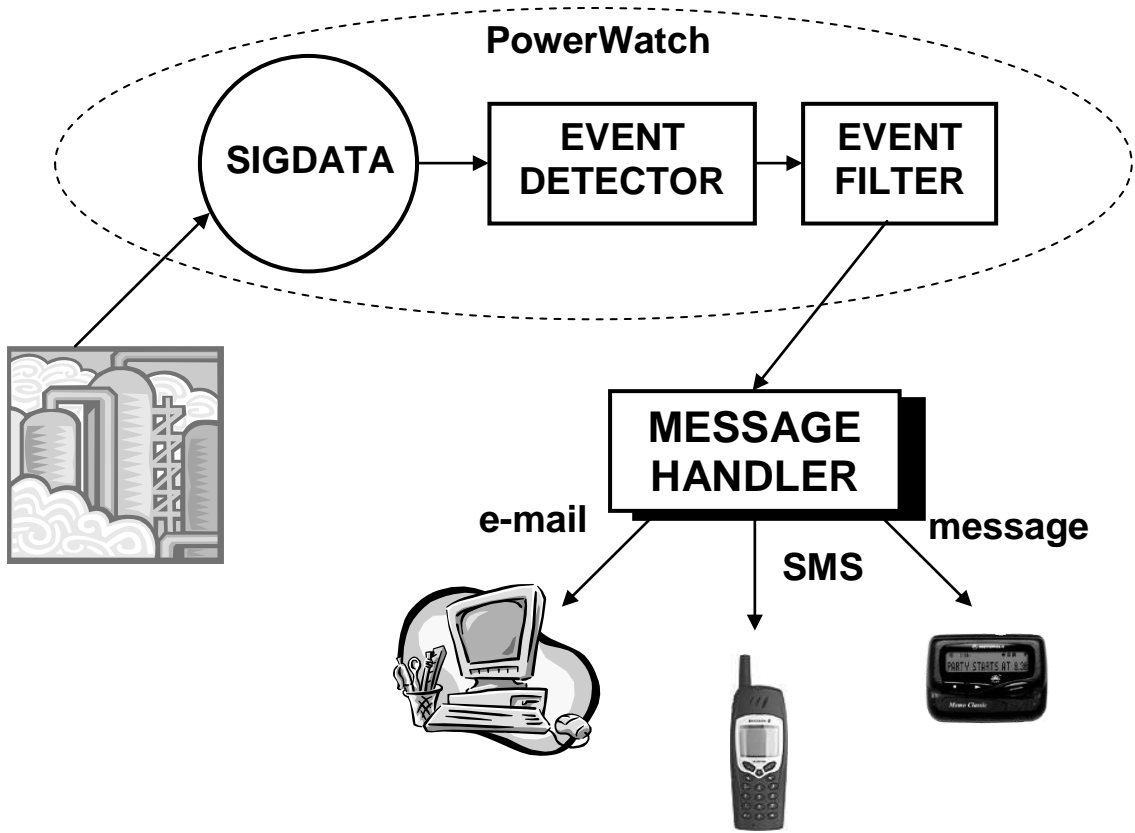

**Fig. 2 The idea of messaging subsystem**

# I.  CASE STUDY – MESSAGE HANDLER

A very good example of a system in which such philosophy may be applied is a pipelined messaging subsystem in an industrial SCADA system called PowerWatch, used for power plant monitoring [Dasz95]. The idea of the messaging subsystem is presented in Fig. 2. Fig. 3 Presents the data flow in the subsystem. The main element of PowerWatch is the real-time data base **SIGDATA**. It collects analog data from **D_SOURCES** (*DATA SOURCES*) installed on industrial plants. Specialized subsystem called **MESSAGING SUBSYSTEM** is responsible for detection of events and alarms from current values of signals compared to catalogue data. In this version of the system there is no difference between events and alarms. In the modified version (not described in this paper) with two-direction communication, alarms require user acceptance but events do not. The **MESSAGING SUBSYSTEM** consists of **EVENT DETECTOR** which identifies situations that are subject of events and alarms and **EVENT FILTER** which lowers the number of messages (events and alarms) using deadbends, timeouts and equalization techniques.

The element we will deal with is **MESSAGE HANDLER**, which structure is in Fig. 4. All the elements supplying data for **MESSAGE HANDLER** (from **D_SOURCE**s to **EVENT FILTER**) are covered by the **EXT** plug. The subsystem under development (**MESSAGE HANDLER**) is showed in all figures with black shadow.

The pieces of data flow through the pipeline of three modules of **MESSAGE HANDLER**: **MES**, **TRANS** and **CH** (shown as dark-gray shadowed boxes in Fig, 4). The **MES** module is responsible for organization of message flow by message numbering. It has two purposes: distinguishing between distinct messages send to one user and identification inside **MESSAGE HANDLER** (in two-direction version of the system it is also used for the organization of sliding window). The **TRANS** module is responsible for routing: it decides to what user the message should be send and thorough which data channel. The **CH** module represents the data channel management. In general it might be a function of **TRANS**, but the system is prepared to handle multiple channels (for example e-mail, SMS message and pager message). The **MESSAGE HANDLER** cooperates with more than one user which is denoted by index (in this case two channels are used).

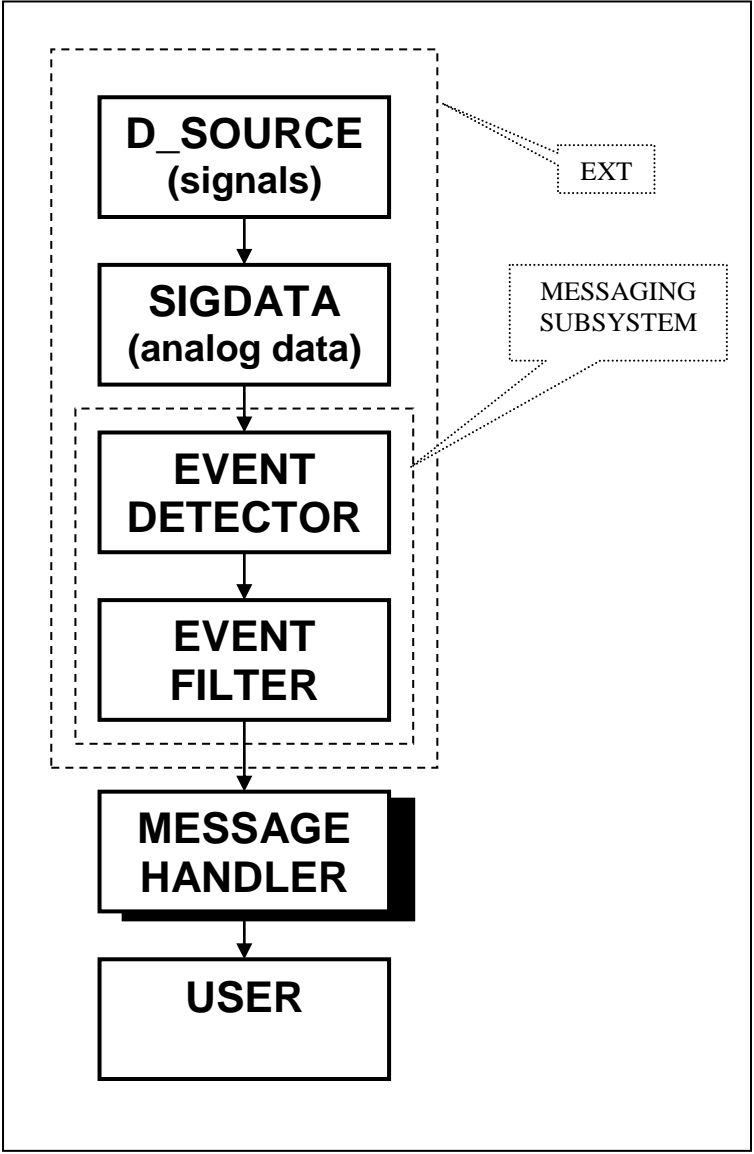

**Fig. 3 General data flow**

The pieces of data flow through the pipeline in one direction: form the source **EXT** downward to one of **USER**s. It is symbolically denoted by a message $M$. The message is equipped with parameters operated by individual modules: **EXT** issues a message with 'alarm/event' parameter ($a\_e$) and type of data contained (*type*). The module **MES** adds a sequence number (*seq*). The module **TRANS** decides to which user a message should be send and adds the parameter *user*. This parameter is consumed by the module **CH** which routes a message to the desired user. Parameters may be implemented in one of two ways:
- attributes of messages in Extended CSM formalism [Krys03], not analyzed by model checking mechanism,
- additional signals passed between modules together with the main signal *M*.

We may choose the way of modeling individually for any of the parameters.

The next step of refinement (Fig. 5) concerns internal structure of modules (identification of sub-modules inside modules) and highlights the protocols used in the system: intra-module communication ① and inter-module communication ②. For example intra-module communication inside the module **MES** is used for generating sequence numbers for messages.

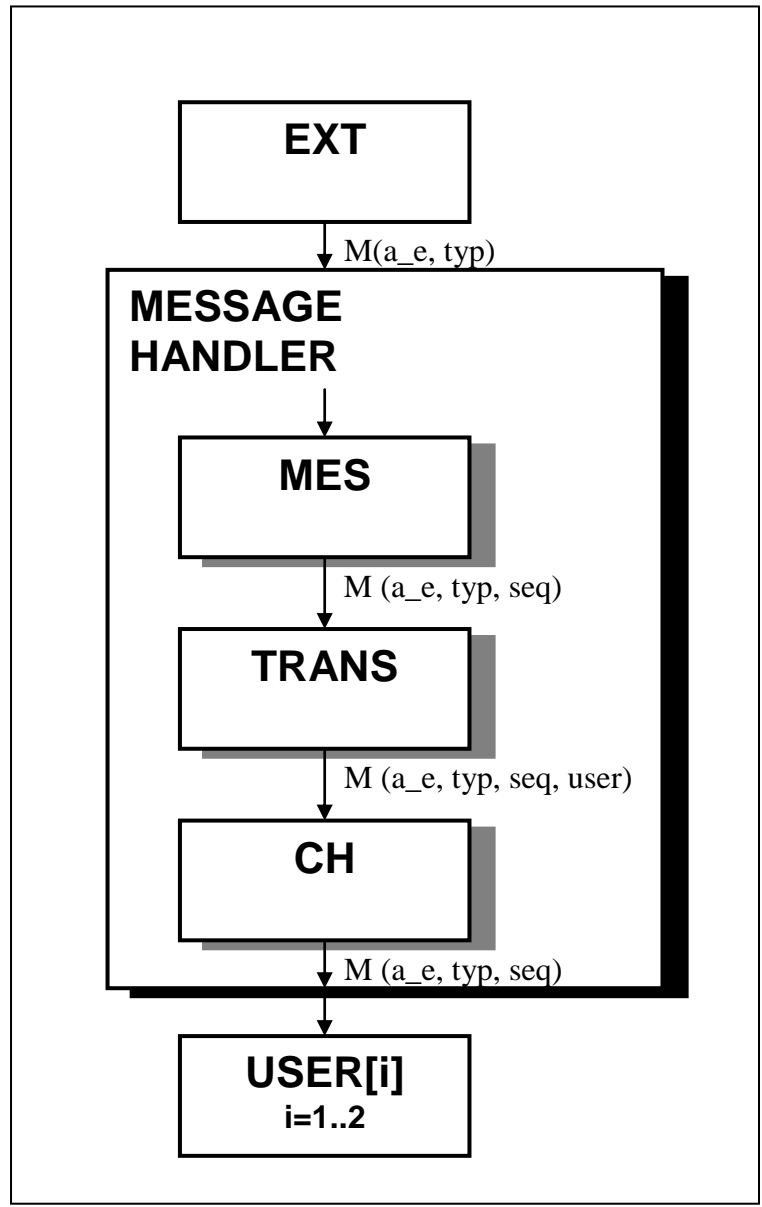

**Fig. 4 Block diagram (one-way system)**

The logical intra-module and inter-module protocols may be standardized across the system, but physical protocols may vary according to i.e. mapping of modules onto servers. For example, **MES** and **TRANS** may be located on a common server, while **CH** on a separate server. In the former case a common memory may be used and a physical protocol may apply signals lasting in time (values of variables). In the latter case signals should be pin-style because they are passed through external medium (a cable). These features require a set of protocols (implemented as communicating CSM automata) which have unified interface to higher-level automata, but various internal structures. This is presented in the example in Fig. 6: a pair of sub-modules **LOGIC** and **GEN**, and additional module implementing the physical protocol. Various physical protocols may be applied on condition that common interface with modules **LOGIC** and **GEN**. The virtual protocol has two phases: **LOGIC** asks **GEN** for parameters by issuing GET_xxx, and gets the reply xxx. For instance in **TRANS** module, the message GET_SEQ is equipped with kind of message (a_e – alarm/event and typ – type).

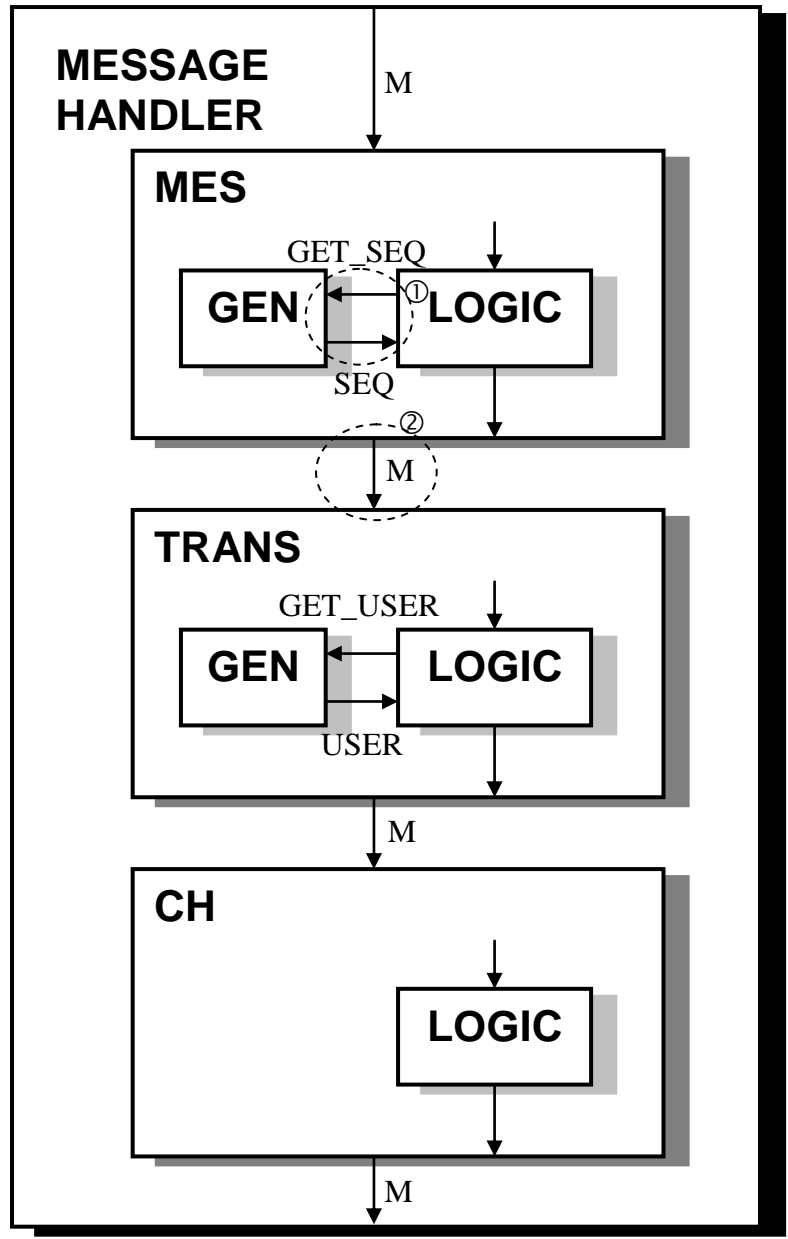

**Fig. 5 Refined block diagram (single-direction system)**

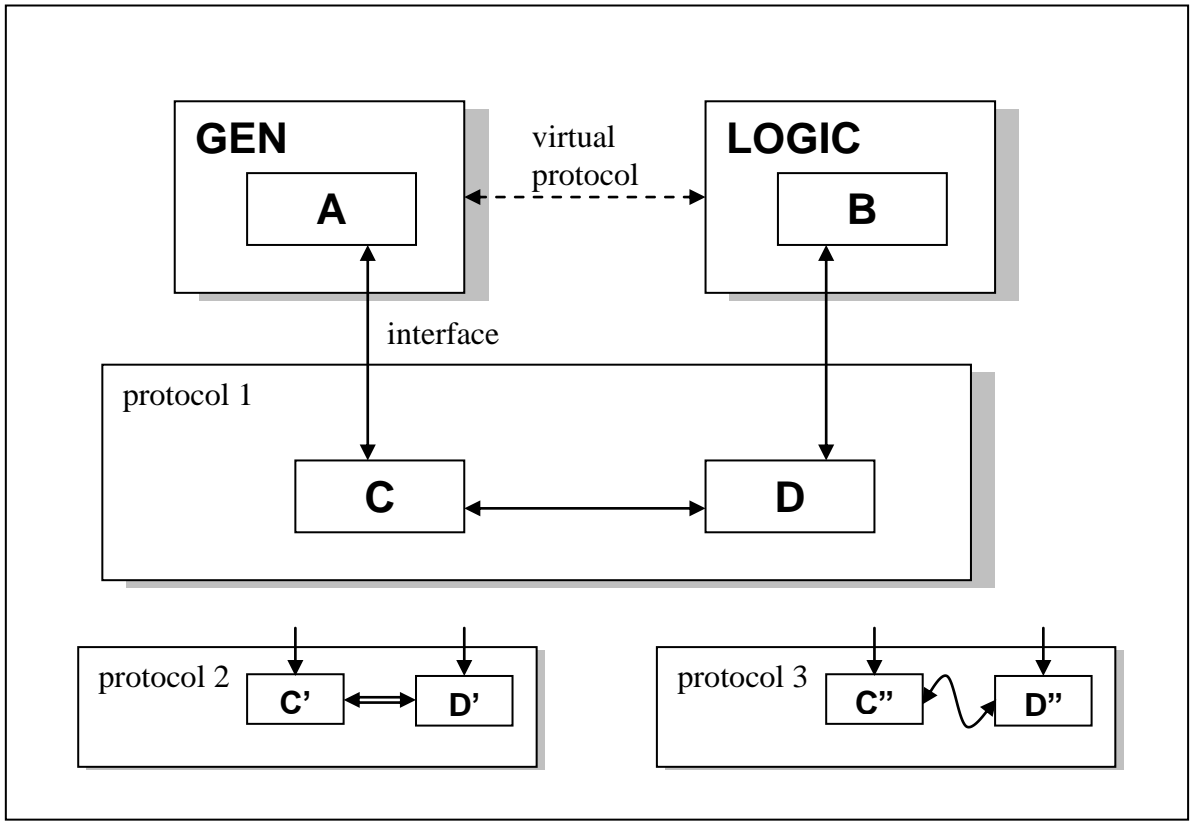

**Fig. 6 Various protocols with common interface to automata A and B**

## II. COMMUNICATION IN THE SYSTEM

INTRA-MODULE COMMUNICATION

The logical protocol inside modules א (**MES** and **TRANS**) consists in simple pair of signals conforming hand-shake: GET_xxx and xxx. We assume the unified physical protocol based on shared memory. The **CALLER** in the protocol is **LOGIC** (initiates the communication), the **CALLEE** is **GEN** (acts as responder). Message sequence in the intra-module communication is shown in Fig 7. The **CALLEE** informs that it is ready to the communication by issuing rdy signal continuously. The **CALLER** upon receiving rdy sends the calling signal get and get reply rep. The protocol is unified, and we have two instances of it in the system: in modules **MES** and **TRANS**. I will show how to use macro-generation technique to replicate the general pattern of the protocol (obtain objects of a class).

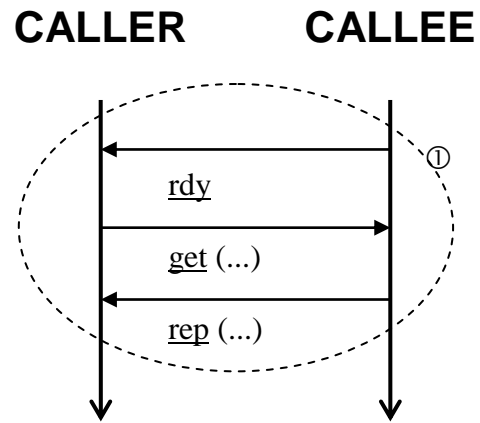

**Fig. 7 Internal protocol ① (general pattern of sequence diagram) between components of the module**

INTER-MODULE COMMUNICATION

The logical protocol consists in only one massage *M* passed from one element of the pipeline to another one. The physical interface between modules (in fact between specific components inside separate modules) is asymmetric – the UPPER MODULE **ACTIVE** (the name suggests that the upper module initiates the communication) calls the LOWER MODULE **PASSIVE** (the responder). In the local (shared memory) communication **ACTIVE** informs that it is ready to the communication by issuing rdy signal continuously. The **PASSIVE** upon receiving rdy sends the calling signal m and gets the acknowledgement ack.

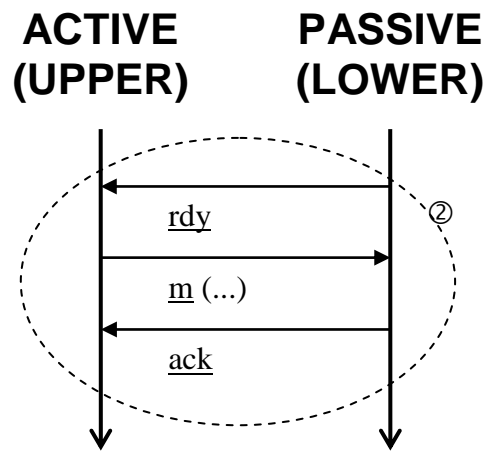

**Fig. 8 Protocol (sequence diagram) between modules ②**

It is only an example of inter-module protocol. It will be designed differently for modules distributed among nodes of the network. In this case the signal rdy cannot be issued continuously, instead the **ACTIVE** may test the state of **PASSIVE** periodically, as long as the signal informing of the ready state of **PASSIVE** is not received. Yet another version (with mutual exclusion mechanism) should be used in two-direction protocol with two **USER**s (both of them may need to initiate communication with **CH** concurrently).

# III. MACROGENERATION TECHNIQUE

Several naming conventions must be used in defining a library-style automatnon and in inseting into a system:
- fixed name of library module to identify it among other library modules (as name of a class),
- fixed name of every automaton inside a library module to identify it among other automata (as name of a subclass),
- formal parameters being signals passed to a module from other automata and reverse (as methods of a class),
- attributes of signals (ECSM attributes) being 'second level parameters'.

Invoking a library module in the system of CSM automata must take care on:
- variable name of an instance of library module (as name of object of given class),
- variable name of an instance of every automaton in library module (as name of object of given subclass),
- actual signals assigned to formal names of signals coming to and from library module,
- actual attributes of signals.

A notation I propose for definition of library module is as follows:
- The identifier of a library module is of the form simple identifier, for example **LIBRARY_MODULE**. Internal signals in the library modules, which should be "made visible" (for example to the model checker or to other tools) are denoted in standard parameter notation: identifiers embraced by parentheses and separated by commas **LIBRARY_MODULE( , , )**.
- The identifier of an automaton is a name, for example **XXX**. It distinguishes the automaton from other automata in the library module. Signals passed to and from automaton are embraced by parentheses and are eapated by commas: **XXX**.
- We will reserve a special character '%' to precede all formal identifiers in library modules. The identifier in whole must fit the naming conventions for automata in COSMA environment [Cwww].
- The name of formal parameter of a module being a signal is simply its identifier preceded by percent: **%x**. It may be either input parameter (used in formulas on transitions of automata inside the module but not generated in any automaton inside the module) or output parameter (generated in a state of an automaton inside the module). Input parameter is preceded by a reserved word **in** while output parameter by **out**.
- The list of parameters (possibly empty) being attributes of a signal is embraced by parentheses and consists of names of attributes preceded by percent: **%x(%y)**.
- Input parameter may be identified with a signal or with a constant value (*0* or *1*).
- Output parameter may be identified with a signal or discarded using reserved word **dummy**.

To sum up, the example full identifiers of a module and of an automaton inside library module are:

   **LIBRARY_MODULE(in %x1(%y1),out %x2(%y2),out %x3)**
   **LIBRARY_MODULE.XXX**

Where **XXX** identifies the automaton, %x1, %x2 and %x3 are signals, *(%y1)* and *(%y2)* are lists of attributes. Formally I will define it in BNF:

```
<digit> ::= 0|1|2|3|4|5|6|7|8|9
<letter> ::= A|…|Z|a|…|z
<underscore> ::= _
<let_dig_un> ::= <letter> | <digit> | <underscore> |
        <letter> <let_dig_un> |
        <digit> <let_dig_un> |
        <underscore> <let_dig_un> |
<identifier> ::= <letter> | <letetr> <let_dig_un>
<qualifier> ::= in | out
<constant_value> ::= _0 | _1
<dummy_parameter> ::= dummy

<formal_attrib_id> ::= %<identifier>
<formal_attrib_list> ::= (<formal_attrib_id>)
<formal_signal_id> ::= <qualifier> %<identifier>
<formal_signal_reference> ::= <formal_signal_id> |
        <formal_signal_id> <formal_attrib_list>
<formal_signal_sequence> ::= <formal_signal_reference> |
        <formal_signal_reference>, <formal_signal_sequence>
<formal_signal_list> ::= (<formal_signal_sequence>)
<auto_id> ::= <identifier>
<formal_module_id> ::= <identifier>
<formal_module_header> ::= <formal_module_id> |
        <formal_module_id> <formal_signal_list>
<formal_automaton_header> ::= <formal_module_id>.<auto_id>

<actual_attrib_id> ::= <identifier>
<actual_attrib_sequence> ::= <actual_attrib_id> |
        <actual_attrib_id>, <actual_attrib_sequence>
<actual_attrib_list> ::= (<actual_attrib_sequence>)
<actual_signal_id> ::= <identifier>
<actual_signal_reference> ::= <actual_signal_id> |
        <actual_signal_id> <actual_attrib_list> |
        <constant_value> | <dummy_parameter>
<actual_signal_sequence> ::= <actual_signal_reference> |
        <actual_signal_reference>, <actual_signal_sequence>
<actual_signal_list> ::= (<actual_signal_sequence>)
<actual_module_id> ::=
        <identifier>: <formal_module_id> |
        <identifier> <actual_signal_list>:
           <formal_module_id>
<actual_automaton_header> ::= <actual_module_id>.<auto_id>
```

The semantic constraint for formal definition are:
- formal identifiers must be unique (i.e. every identifier may be used exactly once in a list of formal parameters and their attributes);
- actual parameters being signals must be assigned to formal parameters uniquely (no signal may be assigned to more than one formal parameter);
- the same concerns identifiers of attributes;
- actual input parameters must be generated outside library module;

- every formal input parameter must be used in at least one formula on a transition inside library module;
- no formal input parameter may be generated in any state inside library module;
- actual output parameters may be used in formulas on transitions outside library module and/or treated as signals for "external world";
- every formal output parameter must be generated inside library module.

For example, for the library module mentioned above and conisting of the only automaton ***A***, its actual invocation may have the form:

***LM(*** s1*(a11,a12)*,s2*(a2)*,s3***):LIBRARY_MODULE**
***AUT:LM.A***

## INTRA-MODULE COMMUNICATION
## (LOCAL) – LIBRARY MODULE
## X(in %do, out %end, out %m(%y1), out %ack(%y2))

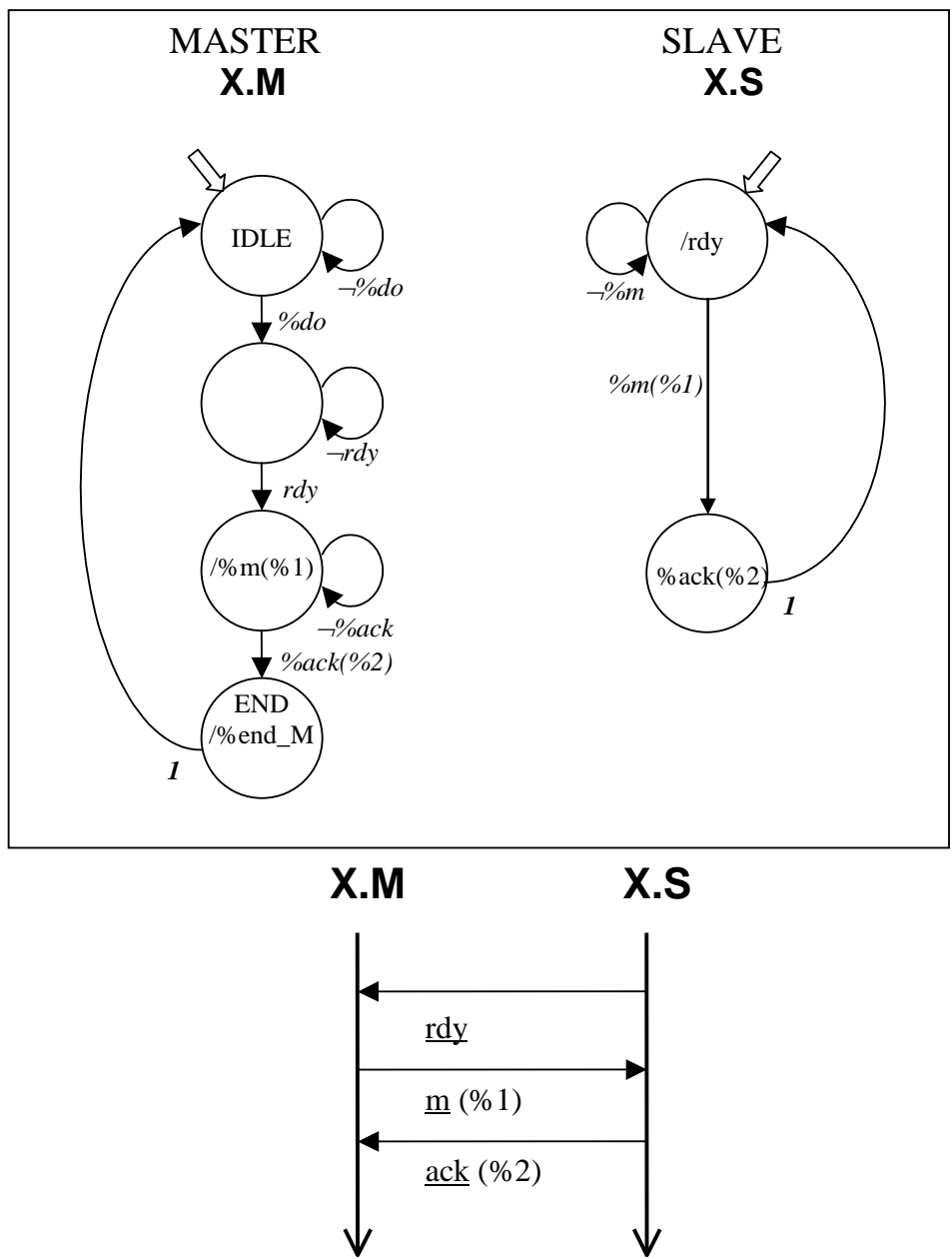

**Fig. 9 Communication between
components inside a module**

where ***LM(**s1(a11,a12),s2(a2),s3**)*** is an instance (object) of the library module, and ***AUT*** is an instance of its component automaton ***A***. While inserting the module into a system, the design environment must:
- create a component automaton for every automaton icluded in the library module,
- assign actual signals and attributes specified in invocation to signals being formal parametres in library module.

In our example, the automaton ***AUT*** is created as an instance of automaton ***LIBRARY_MODULE.A***, signal s1 with attributes *a11* and *a12* is assigned to parameter **%x1*(%y1)*,** and so on.

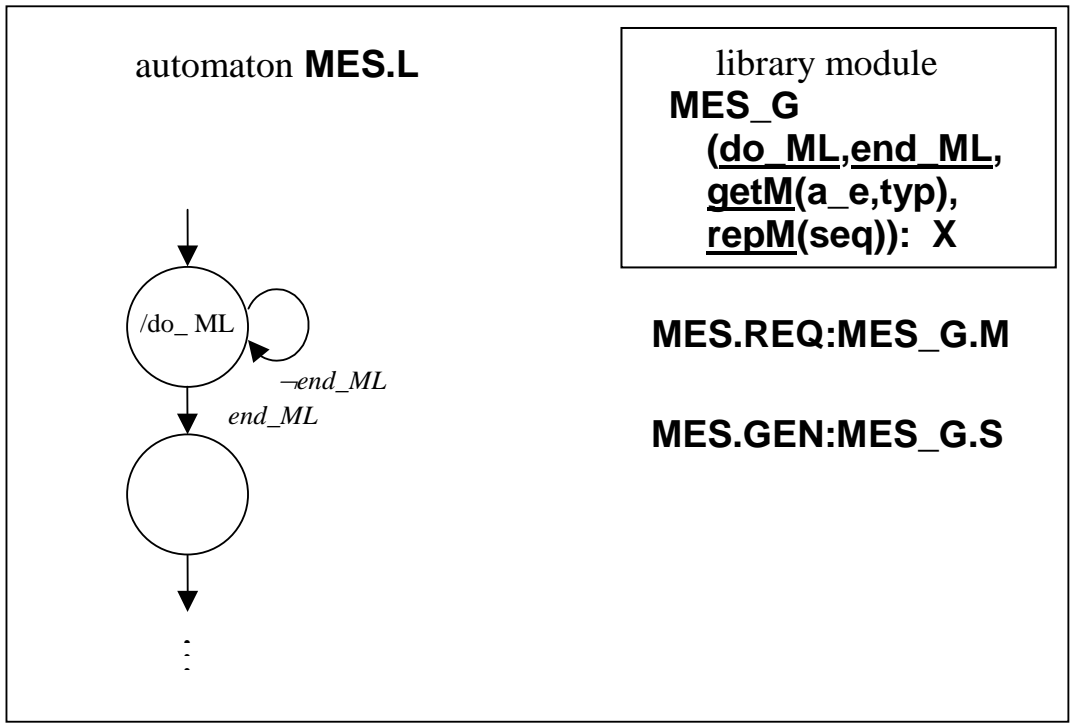

**Fig. 10 Submodule *LOGIC* in module *MES***

## IV. EXAMPLES OF MACROGENERATION

### IV.1. A library module used many times

Let us prepare a library module ***X*** responsible for intra-module communication ① between submodules ***LOGIC*** and ***GEN*** inside modules ***MES*** and ***TRANS***. Assume that the communication is performed locally using shared memoty, i.e. the signals may last in time (impelemnted as values of variables). The structure of the library module is presented in Fig 9. It consists of two automata: MASTER ***X.M*** (it commissions some work to do) and SLAVE ***X.S*** (it does the work). In ***MES*** module, SLAVE allocates a sequence number for a message. In ***TRANS***, SLAVE decides to what user a message will be send. The **virtual protocol** (shown in Fig. 5) consists of two signals: the call m with parameter list *%y1* and the reply ack with return parameter list *%y2*. **Physical protocol** contains additional signal rdy which informs MASTER that SLAVE is ready to accept a call. This signal is internal to the library module, and the system must assign some unique name to it while creating actual instance of a library module.

Both automata (MASTER and SLAVE) are called from higher level automata by means of a structure called **section** [Mieśyy]: section *Sec* is invoked by a signal do_Sec to *Sec* and finished by signal end_Sec from *Sec*. In is a unified **interface** between automata implementing the interface and higher level automata. The structure of module ***MES*** using internal library module ***X*** is presented in Fig. 10.

The library module ***X*** is applied twice in the system: in module ***MES*** and in module ***TRANS***. The instances of ***X*** are: ***MES_G*** and ***TRANS_G*** (***GENERATORS*** for ***MES*** and ***TRANS***). Declarations of instances library module and its automata (actual automata) are:

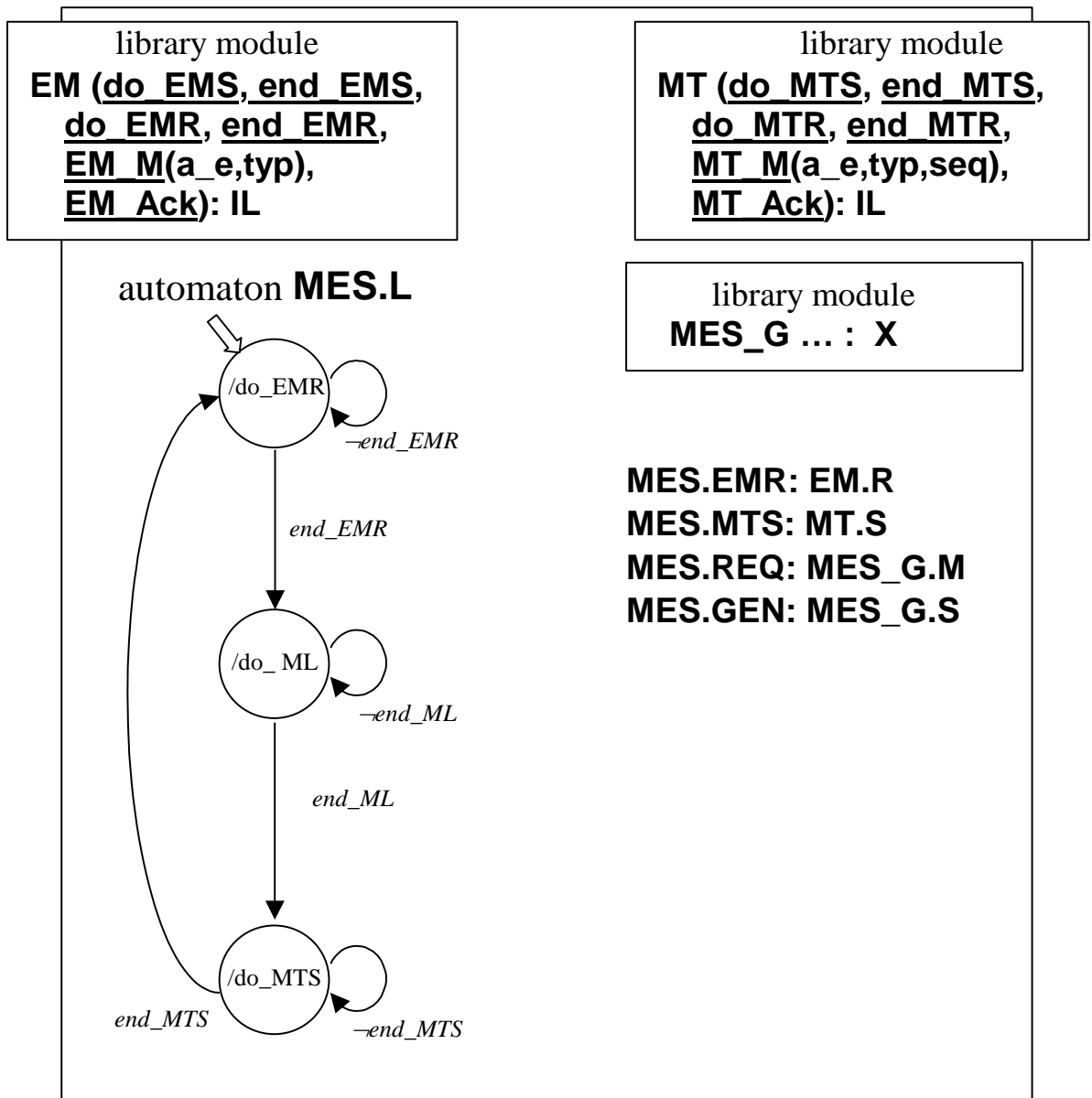

**Fig. 11 Submodule *LOGIC* in module *MES***

*MES_G(<u>do_ML</u>,<u>end_ML</u>, <u>getM</u>(a_e,typ), <u>repM</u>(seq)): X*
*MES.REQ:MES_G.M*
*MES.GEN:MES_G.S*
*TRANS_G(<u>do_TL</u>,<u>end_TL</u>, <u>getT</u>(a_e,typ), <u>repT</u>(user)): X*
*MES.REQ:MES_G.M*
*MES.GEN:MES_G.S*

### IV.2. A set of library modules used alternatively

Inter-module link ② may be organized in various ways, depending on the media of communication. It may be performed locally, as in a case of intra-module link, or remotely in a case of distributed system. In the latter case, signals cannot last in time: they can be issued as messages instantly. One solution is sending a signal periodically until it is accepted; the other way is remembering information of the arrival of the signal until it is used. We prepare three versions of inter-module protocol: local, remote with periodical sending and remote with remembering buffer.

The structure of the submodule **MES.LOGIC** is presented in Fig. 11. Three submodules are used: Completely internal generator **MES_G** (described in section IV.1), receiving part of module **EM** (responsible for communication between **EXT** and **MES**) and sending part of module **MT** (responsible for communication between **MES** and **TRANS**). These modules stick out of the frame containing **MES.LOGIC** to show symbolically that parts of these library modules belong to other modules. Note that both **EM** and **MT** are instances of the same library module **IL**.

Library module **IL** for local (not distributed) communication is quite similar to the library module **X** for intra-module communication. It is presented in Fig. 12. Fig. 13 contains a library module **ID** for distributed communication with periodic issuing of signal rdy. Fig. 14 presents a library module **IB** for distributed communication with a buffer remembering arrival of signal rdy. All three library modules have common interface to higher level automata (compatible to the principles of calling **sections** presented in [Mieś99]) and may be applied alternatively. For example, the type of module MT may be changed from **IL** to **ID** or **IB** without any another changes in **MES.LOGIC** because of unified interface of these three library modules. The situation is similar to that presented in Fig. 6.

## INETR-MODULE COMMUNICATION (LOCAL) – LIBRARY MODULE
## IL(in %do_S, out %end_S, in %do_R, out %end_R, out %m(%y1), out %ack)

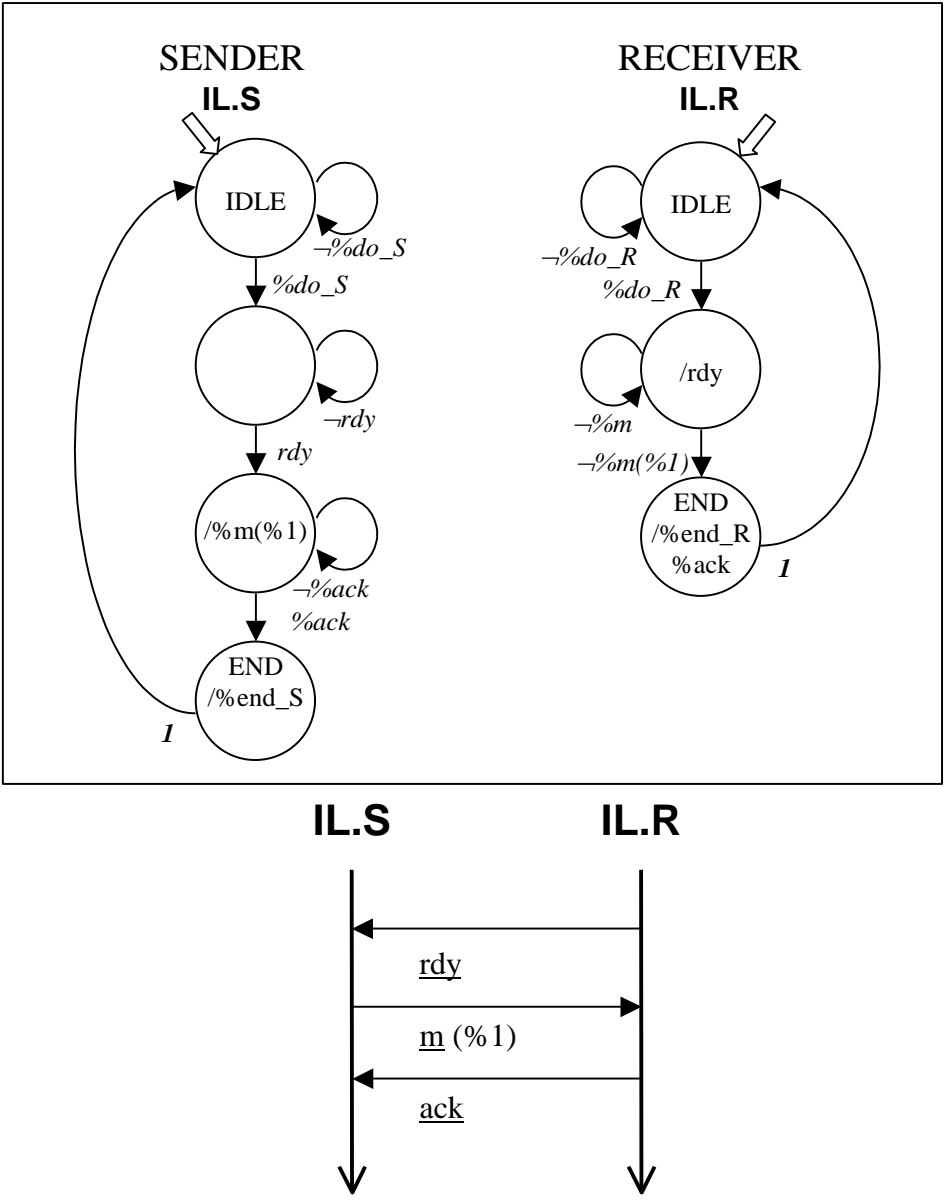

**Fig. 12 Communication between submodules of distinct modules (local)**

**INTER MODULE COMMUNICATION
(DISTRIBUTED) – LIBRARY MODULE
ID(in %do_S, out %end_S, in %do_R,
out %end_R, out %m(%y1), out %ack)**

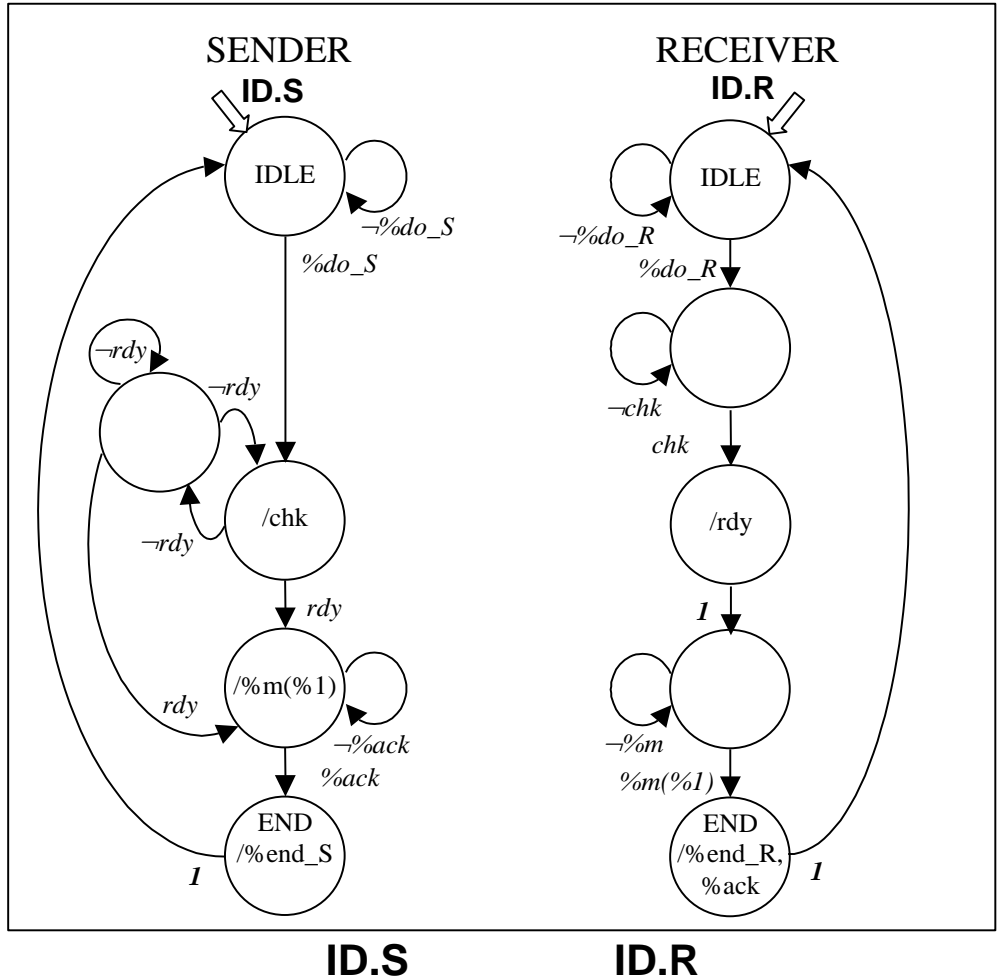

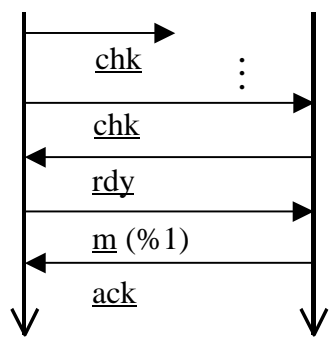

**Fig. 13 Communication between
submodules of distinct modules
(distributed)**

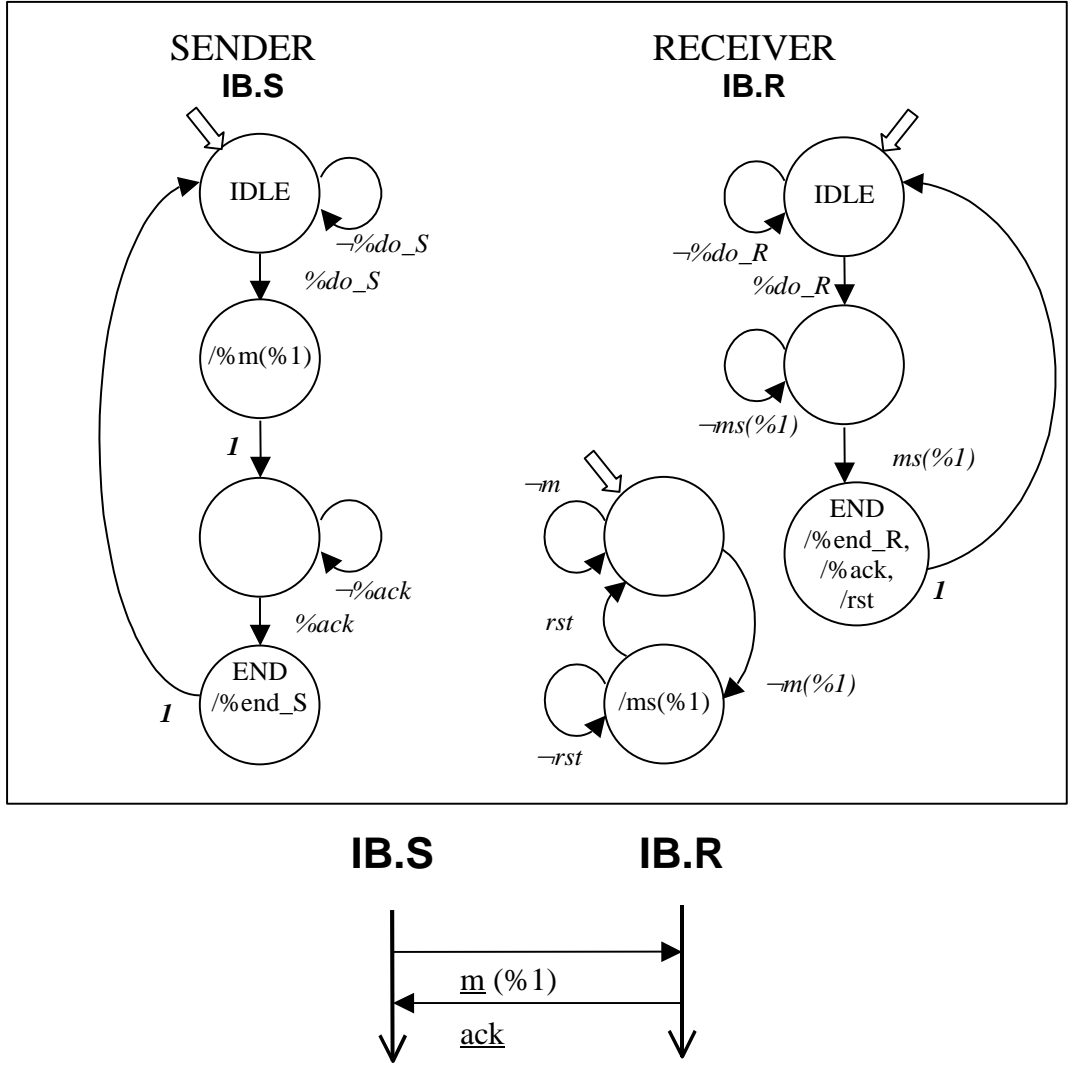

**Fig. 13 Communication between submodules of distinct modules (distributed+buffer)**

## V.  PARAMETERIZED MACROGENERATION

In many cases there is need for an automaton with number of states specified by a parameter. Examples of such automata are counter to *N* and arbiter between *N* requestors. Naturally, such automata cannot be shown precisely as pictures (see Fig. 14; $\varepsilon(s_0,…,s_i)$ is a shortcut of $\neg s_0 * … * \neg s_i$).

I propose to represent parameterized library module by parameterized header, parameterized set of states and parameterized transition table. The example is:

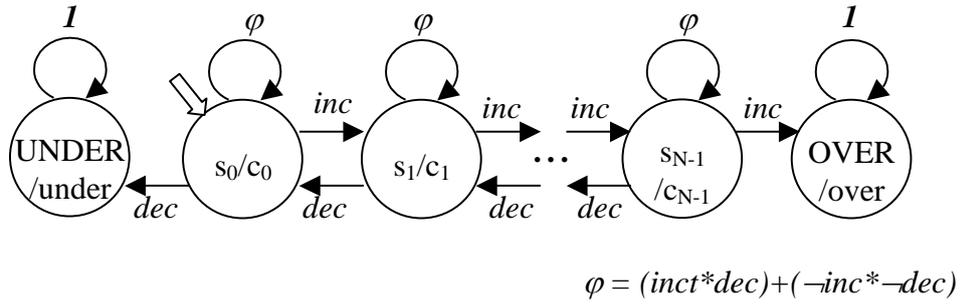

$\varphi = (inc*dec)+(\neg inc*\neg dec)$

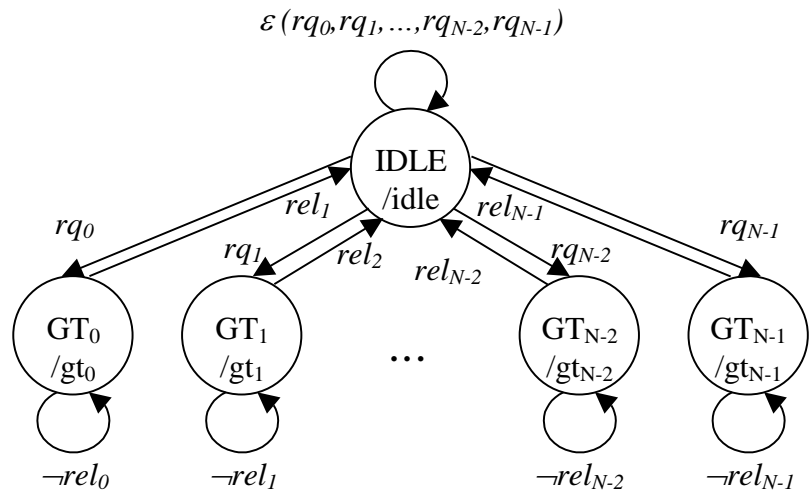

**Fig. 14 Counter to N and arbiter between N requesters**

***LIBRARY_MODULE(%[M],%[N],in <u>%set[1..M]</u>,out <u>%get[0..N-1]</u>)***

In BNF:

```
<expression> ::= <identifier> | <identifier>+<number> |
         <identifier>-<number> | <number>
<range> ::= <expression>..<expression>
<range_list> ::= <range> | <range>, <range_list>
<module_parameter> ::= %'['<identifier>']'
<formal_signal_id> ::= %<identifier>
<formal_signal_reference> ::= <formal_signal_id> |
         <formal_signal_id> <formal_attrib_list>
<formal_signal_vector> ::= %<identifier>'['<range_list>']'
<formal_signal> ::= <formal_signal_reference> |
         <formal_signal_vector>
<formal_reference> ::= <formal_signal> |
         <module_parameter>
```

```
<formal_sequence> ::= <formal_reference> |
        <formal_reference>, <formal_sequence>
<formal_list> ::= (<formal_sequence>)
```

<formal_list> replaces <formal_signal_list> in all further definitions.

```
<index_range> ::= '['<identifier>=<expression>']' |
        '['<identifier>=<range_list>']'
<index_range_list> ::= <index_range> |
        <index_range> <index_range>
<index> ::= <identifier>
<index_list> ::= <identifier> | <identifier>,<index_list>
<indices> ::= '['<index_list>']'

<signal> ::= <identifier> | <identifier> <indices>
<signal_list> ::= <signal> | <signal>,<signal_list>
<state> ::= <identifier> | <identifier> <indices> |
        <identifier> / <signal_list> |
        <identifier> <indices> / <signal_list>

<state_list> ::= <state> | <state>,<state_list>
<states> ::= [<index_range_list>] (<state_list>)
<expression_list> ::= <expression> |
        <expression>,<expression_list>
<initial_state> ::= init <identifier> |
        init <identifier>'['<expression_list>']'

<transition_range> ::= '['<identifier>=<expression>']' |
        '['<identifier>=<range_list>']'
<transition_range_list> ::= <transition_range> |
        <transition_range> <tramsition_range>
<state_index> ::= <identifier> | <expression>
<state_indices> ::= <state_index> |
        <state_index>,<state_indices>
<state_reference> ::= <identifier> |
        <identifier>'['<state_indices>']'
<formula> ::= ...(may contain signals with indices)
<transition> ::= [<transition_range_list>] <state_reference>
        '--{' <formula> '}-->' <state_reference>
```

The counter to N from Fig. 14 is has a following form of library module:

**COUNTER(%[N], in <u>inc</u>, in <u>dec</u>, out <u>under</u>, out <u>over</u>, out <u>c</u>[0..N-1])**
**COUNTER.AUTOMATON**
*[I=0..N-1] (UNDER/under, s[I]/c[I], OVER/over)*
*UNDER --{ 1 }--> UNDER*
*s[0] --{ dec*¬inc }--> UNDER*
*[I=0..N-2] s[I] --{ inc*¬dec }--> s[i+1]*
*s[N-1] --{ inc*¬dec }--> OVER*
*[I=1..N-1] s[i] --{ dec*¬inc }--> s[i-1]*

OVER --{ 1 }--> OVER
[i=0..N-1] s[I] --{ inc*dec + ¬inc*¬dec }--> s[i]

The second version of the counter reaches the state ERROR when signals inc and dec are issued simultaneously:

**NEW_COUNTER(%[N], in inc, in dec, out under, out over, out err, out c[0..N-1])**
**NEW_COUNTER.AUTOMATON**
[I=0..N-1] (UNDER/under, s[I]/c[I], OVER/over, ERR/err)
ERR--{ 1 }--> ERR
UNDER --{ 1 }--> UNDER
s[0] --{ dec*¬inc }--> UNDER
[I=0..N-2] s[I] --{ inc*¬dec }--> s[i+1]
s[N-1] --{ inc*¬dec }--> OVER
[I=1..N-1] s[i] --{ dec*¬inc }--> s[i-1]
OVER --{ 1 }--> OVER
[i=0..N-1] s[I] --{ ¬inc*¬dec }--> s[i]
[i=0..N-1] s[I] --{ inc*dec }--> ERR

Another version makes it possible to set the counter to specific value:

**SET_COUNTER**
**(%[N], in inc, in dec, out under, out over, in set[0..N-1], out c[0..N-1])**
**SET_COUNTER.AUTOMATON**
[I=0..N-1] (UNDER/under, s[I]/c[I], OVER/over)
ERR--{ 1 }--> ERR
UNDER --{ 1 }--> UNDER
s[0] --{ dec*¬inc }--> UNDER
[I=0..N-2] s[I] --{ inc*¬dec }--> s[i+1]
s[N-1] --{ inc*¬dec }--> OVER
[I=1..N-1] s[i] --{ dec*¬inc }--> s[i-1]
OVER --{ 1 }--> OVER
[i=0..N-1] s[I] --{ inc*dec + ¬inc*¬dec }--> s[i]
[i=0..N-1][j=0..N-1] s[I] --{ set[j] }--> s[j]

The two indices in the last transition generate a Cartesian product. Note the inndeterminism: if both signals inc and $set_j$ are issued, the counter may increment or set to *j*. Similar situation concerns two set signals ($set_i$ and $set_j$) issued simultaneously. The indeterminism is difficult do avoid without the symbol $\varepsilon$, set of indexed signals and inequality of indices. I propose to use additional symbols for shortcuts of formulas:

$single(c_0,...,c_i) = c_0 * \neg c_1 * ... * \neg c_i + \neg c_0 * c_1 * ... * \neg c_i + ... + \neg c_0 * \neg c_1 * ... * c_i$
$single_j(c_0,...,c_i) = \neg c_0 * \neg c_1 * ... * c_j * ... * \neg c_i$
$any\ (c_0,...,c_i) = c_0 + c_1 + ... + \neg c_i$
$all\ (c_0,...,c_i) = c_0 * c_1 * ... * \neg c_i$

**DETERMINISTIC_COUNTER**
**(%[N], in inc, in dec, out under, out over, out err, in set[0..N-1], out c[0..N-1])**
**DETERMINISTIC_COUNTER.AUTOMATON**
[I=0..N-1] (UNDER/under, s[I]/c[I], OVER/over, ERR/err)

*ERR --{ 1 }--> ERR*
*UNDER --{ 1 }--> UNDER*
*s[0] --{ dec*¬inc*ε(s[I=0..N-1]) }--> UNDER*
*[I=0..N-2] s[I] --{ inc*¬dec*ε(set[I=0..N-1]) }--> s[i+1]*
*s[N-1] --{ inc*¬dec*ε(set[I=0..N-1]) }--> OVER*
*[I=1..N-1] s[i] --{ dec*¬inc*ε(set[I=0..N-1]) }--> s[i-1]*
*OVER --{ 1 }--> OVER*
*[i=0..N-1] s[I] --{ ¬inc*¬dec*ε(set[I=0..N-1]) }--> s[i]*
*[i=0..N-1][j=0..N-1] s[I] --{ ¬inc*¬dec*single[j](set[k=0..N-1]) }--> s[j]*
*[i=0..N-1] s[I] --{ (inc+dec)*any(set[I=0..N-1]) }--> ERR*
*[i=0..N-1][j=0..N-1][i≠j] s[I] --{ set[I]*set[j] }--> ERR*
*[i=0..N-1] s[I] --{ inc*dec }--> ERR*

The arbiter between N requesters is:

**ARBITER (%[N], in rq[0..N-1], in rel[0..N-1], out gt[0..N-1], out idle)**
**ARBITER.AUTOMATON**
*[I=0..N-1] (IDLE/idle, GT[I]/gt[I])*
*IDLE --{ ε(rq[I=0..N-1]) }--> IDLE*
*[I=0..N-1] IDLE --{ rq[I] }--> GT[I]*
*[I=0..N-1] GT[i] --{ ¬rel[I] }--> GT[I]*
*[I=0..N-1] GT[I] --{ rel[I]* }--> IDLE*

Note the desired indeterminism: $\underline{rq}_i$ is used instead of *single*($\underline{rq}_i$) on purpose.

The second version can "switch" between grants ($GT_i$) if request ($\underline{rq}_j$) is issued concurrently with release ($rel_i$):

**SWITCH (%[N], in rq[0..N-1], in rel[0..N-1], out gt[0..N-1], out idle)**
**SWITCH.AUTOMATON**
*[I=0..N-1] (IDLE/idle, GT[I]/gt[I])*
*IDLE --{ ε(rq[I=0..N-1]) }--> IDLE*
*[I=0..N-1] IDLE --{ rq[I] }--> GT[I]*
*[I=0..N-1] GT[i] --{ ¬rel[I] }--> GT[I]*
*[I=0..N-1] GT[I] --{ rel[I]*ε(rq[j=1..N-1]) }--> IDLE*
*[I=0..N-1][j=0..N-1] GT[I] --{ rel[I]*rq[j] }--> GT[j]*

To sum up, the syntax of library modules is as follows (semantic constraints are given in italics):

```
<digit> ::= 0|1|2|3|4|5|6|7|8|9
<letter> ::= A|…|Z|a|…|z
<underscore> ::= _
<let_dig_un> ::= <letter> | <digit> | <underscore> |
        <letter> <let_dig_un> |
        <digit> <let_dig_un> |
        <underscore> <let_dig_un> |
<number> ::= <digit> | <digit> <number>
<identifier> ::= <letter> | <letetr> <let_dig_un>
<qualifier> ::= in | out
```

```
<constant_value> ::= _0 | _1

<expression> ::= <identifier> | <identifier>+<number> |
        <identifier>-<number> | <number>
```
*identifier must be a module parameter or an index identifier*
```
<range> ::= <expression>..<expression>
<range_list> ::= <range> | <range>, <range_list>

<module_parameter> ::= %'['<identifier>']'
<module_parameter_list> ::= <module_parameter> |
        <module_parameter>,<module_parameter_list>
```
*module parameters must use unique identifiers*
```
<formal_attrib_id> ::= %<identifier>
```
*formal attrib id identifier must use unique identifier*
```
<formal_attrib_list> ::= (<formal_attrib_id>)
<formal_signal_id> ::= <qualifier> %<identifier>
<formal_signal_reference> ::= <formal_signal_id> |
        <formal_signal_id> <formal_attrib_list>
```
*formal signal ids must use unique identifiers*
```
<formal_signal_vector> ::= %<identifier>'['<range_list>']'
<formal_signal> ::= <formal_signal_reference> |
        <formal_signal_vector>
<formal_signal_list> ::= <formal_signal> |
        <formal_signal>,<formal_signal_list>
<formal_sequence> ::= <module_parameter_list> |
        <formal_signal_list>|
        <module_parameter_list>,<formal_signal_list>
<formal_list> ::= (<formal_sequence>)
<formal_module_id> ::= <identifier>
<formal_module_header> ::= <formal_module_id> |
        <formal_module_id> <formal_list>
<auto_id> ::= <identifier>
<formal_automaton_header> ::= <formal_module_id>.<auto_id>
```
*distinct automata must use unique identifiers in library module*

```
<index_range> ::= '['<identifier>=<expression>']' |
        '['<identifier>=<range_list>']'
```
*1. index identifier is local to a definition of state vector or a transition*
*2. index identifiers must use unique identifiers in single definition*
```
<inequality> ::= <identifier>/=<expression>
```
*identifier must be an index identifier defined prior to inequality*
```
<range_element> ::= <index_range> | <inequality>
<index_range_list> ::= <range_element> |
        <range_element> <index_range_list>
<index> ::= <expression>
<index_list> ::= <index> | <index>,<index_list>
<indices> ::= '['<index_list>']'

<signal> ::= <identifier> | <identifier> <indices>
```

1. *simple identifier must be a simple formal parameter of a library module or a signal internal to the library module*
2. *indexed identifier must be a vector formal parameter of a library module or a signal inernel to the library module, generated in vectored state of a library automaton*

```
<signal_list> ::= <signal> | <signal>,<signal_list>
<state> ::= <identifier> | <identifier> <indices> |
        <identifier> / <signal_list> |
        <identifier> <indices> / <signal_list>

<state_list> ::= <state> | <state>,<state_list>
<states> ::= [<index_range_list>] (<state_list>)
<initial_state> ::= init <identifier> |
        init <identifier><indices>

<transition_range> ::= '['<identifier>=<expression>']' |
        '['<identifier>=<range_list>']'
<transition_range_list> ::= <transition_range> |
        <transition_range> <tramsition_range>
<state_reference> ::= <identifier> | <identifier><indices>

<signal_element> ::= <signal> | <signal><index_range>
<signal_set> ::= <signal_element> |
        <signal_element>,<signal_set>
<any> ::= any(<signal_set>)
<all> ::= all(<signal_set>)
<no_signal> ::= ε(<signal_set>)
<single> ::= single(<signal_set>)
<single_n> ::= single[<indices>](<signal_set>)

<factor> ::= <signal>[<indices>] | <any> | <all> |
        <no_signal> | <signgle> | <single_n> | ¬<factor> |
        (<formula>)
<term> ::= <factor>*<factor> | <factor>*<term>
<formula> ::= <term>+<term> | <term>+<formula>
<transition> ::= [<transition_range_list>] <state_reference>
        '--{' <formula> '}-->' <state_reference>

<dummy_parameter> ::= dummy
<actual_attrib_id> ::= <identifier>
<actual_attrib_sequence> ::= <actual_attrib_id> |
        <actual_attrib_id>, <actual_attrib_sequence>
<actual_attrib_list> ::= (<actual_attrib_sequence>)
<actual_signal_id> ::= <identifier>
<actual_signal_reference> ::= <actual_signal_id> |
        <actual_signal_id> <actual_attrib_list> |
        <constant_value> | <dummy_parameter>
<actual_module_parameter> ::= <number>
<actual_parameter> ::= <actual_signal_reference> |
        <actual_module_parameter>
```

```
<actual_sequence> ::= <actual_parameter> |
            <actual_parameter>, <actual_sequence>
<actual_signal_list> ::= (<actual_signal_sequence>)
```
1. *actual parameters must be compatible with formal parameters (number of parameters, module parameter/signal, existence/lack of attributes)*
2. *input parameter may be assigned only to a signal generated outside library module or a constant*
3. *output parameter may be assigned only to a signal not generated outside library module (used in transitions outside the module) or rejected*
4. **note** *that vectored signals cannot be used as actual parameters: only individual signals are allowed*

```
<actual_module_id> ::=
            <identifier>: <formal_module_id> |
            <identifier> <actual_signal_list>:
               <formal_module_id>
<actual_automaton_header> ::= <actual_module_id>.<auto_id>
```

# VI. CONCLUSIONS

In the report a useful concept of macrogeneration of automata in CSM systems is presented. The idea allows the designer to use many instances of a single module (automata stored in a library as a cooperating set), or to use one module from many ones with unified interface. This is especially useful in modeling of protocols, which was illustrated in the present report. A notation similar to classes and methods is proposed. It consists of definition of a library module (as a class) and creating an instance of a module (like declaration of an object of given class). The notation is a proposition of syntax for macrogeneration in COSMA environment [Cwww].

The notation allows also for parameterized macrogeneration used to specify a size of actual automata via parameter. The typical examples of such cases where macrogeneration is useful are:
- buffer of unknown capacity (parameter $N$ – the size of buffer),
- arbiter switching access to a resource between users (parameter $N$ – the number of users).

Further research will concern simplified notation for variables (for example a simple $N$-valued variable may be treated a counter to $N$) and synchronization structures.

# REFERENCES

[Cwww]      http://www.ii.pw.edu.pl/cosma
[Dasz95]    Daszczuk W. B., Mieścicki J., Lewandowski J., Świrski K., 1995, "Integration of Digital Plant Control System with Company's Management Network", in *Proc. of the Conf. "Research Problems in Heat Energy Industry"*, Warsaw, Dec. 5-8, vol 1, pp. 101-107 (in Polish)
[Dasz01]    Daszczuk W. B., Grabski W., Mieściski J., Wytrębowicz J., 2001, "System Modeling in the COSMA Environment", Proc. Euromicro Symposium on Digital Systems Design - Architectures, Methods and Tools, September 4-6, Warsaw, Poland, pp. 152-157

| | |
|---|---|
| [Krys03] | Krystosik A., "ECSM - Extended Concurrent Stale Machines", Institute of Computer Science, WUT, Research Report 2/2003 |
| [Mieś94] | Mieścicki J., 1994, "Boolean Formulas and the Families of Sets", Bull. Polish Ac. of Sc., Sci. Tech, vol. 42, No 1, 1994 |
| [Mieś99] | Mieścicki J., 1994, "Hierarchical Concurrent State Machines: a Framework for Concurrent Systems' Behavioral Specification and Verification, Institute of Computer Science, WUT, Research Report 13/99 |
| [Mieś03] | Mieścicki J., 2003, "Concurrent State Machines, the formal framework for model-checkable systems", Institute of Computer Science, WUT, Research Report 5/2003 |